\documentclass[11pt,a4paper]{article}

\usepackage{times}
\usepackage{epsfig}
\usepackage{amssymb}
\usepackage{subfigure}
\usepackage{graphicx}
\usepackage{color}
\usepackage{jheppub} 

\begin{document}

\newcommand{\nc}{\newcommand}
\newcommand{\rnc}{\renewcommand}


\makeatletter
\rnc{\theequation}{\thesection.\arabic{equation}}
\@addtoreset{equation}{section}
\makeatother


 \renewcommand{\thefootnote}{\fnsymbol{footnote}}

\newcommand{\tcb}{\textcolor{blue}}
\newcommand{\tcr}{\textcolor{red}}
\newcommand{\tcg}{\textcolor{green}}


\def\be{\begin{eqnarray}}
\def\ee{\end{eqnarray}}
\def\nn{\nonumber\\}


\def\ct{\cite}
\def\la{\label}
\def\eq#1{\eqref{#1}}


\def\a{\alpha}
\def\b{\beta}
\def\g{\gamma}
\def\G{\Gamma}
\def\d{\delta}
\def\D{\Delta}
\def\e{\epsilon}
\def\et{\eta}
\def\ph{\phi}
\def\Ph{\Phi}
\def\ps{\psi}
\def\Ps{\Psi}
\def\k{\kappa}
\def\l{\lambda}
\def\L{\Lambda}
\def\m{\mu}
\def\n{\nu}
\def\th{\theta}
\def\Th{\Theta}
\def\r{\rho}
\def\s{\sigma}
\def\S{\Sigma}
\def\ta{\tau}
\def\o{\omega}
\def\O{\Omega}
\def\pr{\prime}


\def\half{\frac{1}{2}}
\def\goto{\rightarrow}

\def\na{\nabla}
\def\grad{\nabla}
\def\curl{\nabla\times}
\def\div{\nabla\cdot}
\def\pa{\partial}
\def\fr{\frac}

\def\bra{\left\langle}
\def\ket{\right\rangle}
\def\lb{\left[}
\def\lc{\left\{}
\def\ls{\left(}
\def\lp{\left.}
\def\rp{\right.}
\def\rb{\right]}
\def\rc{\right\}}
\def\rs{\right)}

\def\vac#1{\mid #1 \rangle}


\def\td#1{\tilde{#1}}
\def\check{ \maltese {\bf Check!}}


\def\Tr{{\rm Tr}\,}
\def\det{{\rm det}}
\def\text#1{{\rm #1}}


\def\bc#1{\nnindent {\bf $\bullet$ #1} \\ }
\def\ch {$<Check!>$ }
\def\ss {\vspace{1.5cm}}
\def\inf{\infty}

\begin{titlepage}

\hfill\parbox{2cm} { }

\hspace{12cm} \today
 
\vspace{2cm}

\begin{center}
{\Large \bf Correlation functions in expanding universes}

\vskip 1. cm
{Chanyong Park$^{a}$ \footnote{e-mail : cyong21@gist.ac.kr}},
{Hanse Kim$^{a}$ \footnote{e-mail : powerblo@gist.ac.kr}} and 
{Kyungchan Cho$^{a,b}$ \footnote{e-mail : cho615@purdue.edu}} 

\vskip 0.5cm

$^{a}$ {\it  Department of Physics and Photon Science, Gwangju Institute of Science and Technology \\  Gwangju  61005, Korea}\\
$^{b}$ {\it Department of Physics and Astronomy, Purdue University \\ West Lafayette 47907, IN, USA} 
\end{center}

\vskip1.5cm


\centerline{\bf ABSTRACT} \vskip 4mm

By using the braneworld model, we investigate the time evolution of microscopic and macroscopic correlations in expanding universes. To describe the FLRW cosmologies in the holographic setup, we take into account a braneworld moving in the $p$-brane gas geometry, where the radial motion of the braneworld determines the cosmology in the braneworld. We show that the braneworld model reproduces the standard cosmology exactly. In this braneworld model, we investigate the time-dependent mutual information between two disjoint macroscopic subregions and the time-dependent two-point functions in the expanding universes. We find that the mutual information becomes zero when the distance between two subregions is slightly larger than the subsystem size. We also find that it decreases as time and the density of matter increase. However, the microscopic two-point function in the short-distance limit decreases by a power law, whereas it is exponentially suppressed in the long-distance limit due to the screening effect. In addition, we find that the two-point function is also suppressed by a power law with time.

\vspace{1cm}


\end{titlepage}

\renewcommand{\thefootnote}{\arabic{footnote}}
\setcounter{footnote}{0}

\tableofcontents


\section{Introduction}

According to the AdS/CFT correspondence \cite{Maldacena:1997re, Gubser:1998bc, Witten:1998qj, Witten:1998zw}, nonperturbative quantum field theory (QFT) involving all quantum effects can be described by a one-dimensional higher gravity theory. Using this holographic technique in the last two decades, many studies have tried to figure out various nonperturbative features of strongly interacting systems in nuclear and condensed matter physics \cite{Bergman:2007wp, Casalderrey-Solana:2011dxg, Hartnoll:2009ns, Preis:2010cq, Rozali:2007rx, Davis:2008nv, Baggioli:2016rdj, Charmousis_2010, Hong:2007dq, Hashimoto:2009ys, Franco_2010}. Another important feature of holography is that the extra dimension of a bulk geometry is regarded as an energy scale measuring the dual QFT. This implies that a bulk geometry can explain the renormalization group (RG) flow of a dual QFT, which gives us information about the nonperturbative properties of IR physics. Applying the holographic method, in this work, we investigate the time evolution of nonperturbative quantum correlations in various expanding universes.  

Holography provides a fascinating and useful tool for understanding strongly interacting systems. Some important quantities of strongly interacting systems can be realized as simple geometrical objects on the dual gravity side. For example, the $q \bar{q}$-potential and quantum entanglement entropy can be evaluated by a minimal surface extending to the dual geometry \ct{Maldacena:1997re, Rey:1998ik, Herzog:2006ra, Karch:2006pv, Park:2009nb, Ryu:2006ef, Ryu:2006bv, Casini:2008cr, Myers:2010tj, Casini:2011kv, Kim:2016jwu, Park:2015afa}. Moreover, it was also proposed that correlation functions of QFT are calculated by a geodesic length connecting two boundary operators 
\ct{Susskind:1998dq,Balasubramanian:1999zv,Louko:2000tp,Solodukhin:1998ec,DHoker:1998vkc}
\be
\bra O(\ta_1,\vec{x}_1)  O(\ta_2, \vec{x}_2) \ket = e^{- \D \,  L (\ta_1, \vec{x}_1;\ta_2,\vec{x}_2) /R} ,   \la{Formula:TPF}
\ee
where $\D$ and $L (\ta_1, \vec{x}_1;\ta_2,\vec{x}_2) $ indicate a conformal dimension and geodesic length, respectively. It was shown that the holographic proposal \eq{Formula:TPF} could correctly reproduce the known correlation functions of conformal field theory (CFT) in a flat, dS and AdS space \ct{Kim:2023fbr, Chu:2016uwi, Chu:2016pea, Koh:2018rsw}. Moreover, the holographic formula in \eq{Formula:TPF} was exploited to look into quantum correlation in a medium composed of various matters \ct{Park:2022abi, Park:2022mxj, Georgiou:2022ekc, Rodriguez_Gomez_2021, Rodriguez_Gomez_2021_2, Fuertes_2009, Krishna_2021, Volovich_2009}.

The holography is also applied to the cosmological model. In the standard Friedmann-Lema\^{ı}tre-Robertson-Walker (FLRW) cosmology, the Friedmann equation allows various expanding universes as a solution where the rate of expansion is influenced by the equation of state parameter of components filling in the universe. In the holography studies, this FLRW cosmology can be realized in two different ways: the dS boundary \ct{Koh:2020rti, Park:2020xho, Barcelo_2000} and the braneworld models \ct{Randall:1999ee, Randall:1999vf, Kraus:1999it, Park:2000ga, Park:2021wep, Park:2020jio}. For the dS boundary model, we consider a $(d+1)$-dimensional AdS space with a $d$-dimensional dS boundary. Following the AdS/CFT correspondence, the classical gravity in a $(d+1)$-dimensional AdS space can describe nonperturbative quantum physics in an eternally inflating universe. After considering a three-dimensional AdS space as a toy model, time-dependent quantum entanglement entropy in a two-dimensional inflating universe has been investigated \ct{Koh:2020rti, Park:2020xho, Barcelo_2000}. For the dS boundary model, however, it is not easy to describe the FLRW cosmologies expanding by power law because their dual geometries are not known yet. To detour this problem, one can take into account another holographic model called the braneworld model, where the FLRW cosmologies are described by the radial motion of a braneworld in the $p$-brane gas geometry \ct{Park:2021wep, Park:2020jio}. Applying the braneworld model, the leading time evolution of the entanglement entropy in the FLRW cosmology has been studied \ct{Park:2020xho, Park:2021wep, Park:2020jio, Kim:2023fbr}.

By applying the holographic technique in the present work, we investigate time-dependent correlation functions in expanding universes. The correlation functions in cosmology give us important information about the history and structure of our universe, such as the power spectrum and non-Gaussianity \ct{kinney2009tasi}. For expanding universes in the radiation and matter-dominated era, it is not easy to calculate the two-point function on the QFT side because it is defined in a nontrivial curved spacetime. Assuming that one kind of matter with a specific equation of state parameter is uniformly distributed in the universe, it was known that the universe in the standard FLRW cosmology expands by
\be
a(t) \sim t^{\fr{2}{3 (1+w)}} ,
\ee  
where $w$ means an equation of state parameter. After reproducing this scale factor in the braneworld model, in this work, we study microscopic and macroscopic correlation functions. We first consider mutual information describing the entanglement between two disjoint macroscopic subregions. Mutual information may be associated with the collective behavior of microscopic correlation functions. Improving the entanglement entropy calculation in Ref. \ct{Park:2020jio, Park:2021wep}, we further investigate the mutual information in expanding universes. 

We also evaluate the time-dependent two-point functions in various expanding universes. For CFT, the two-point function is suppressed by a power-law due to the scaling symmetry. This is not the case in the medium because the medium breaks the conformal symmetry. A general form of a two-point function in a medium is exponentially suppressed due to the screening effect of the background matter. In the expanding universes, there is an additional effect reducing the two-point function. The expansion of the universe leads to the increase in the physical distance between two operators. Therefore, one expects that the two-point function decreases with time in expanding universes. To clarify the time dependence of two-point functions in expanding universes, we explicitly calculate two-point functions in the braneworld model. To do so, we first show how the braneworld model can reproduce the known two-point function in the inflating universe, which was derived in the dS boundary model \ct{Kim:2023fbr}. After that, we consider a braneworld model describing the other FLRW cosmologies and calculate two-point functions. We find that the two-point function in the late time era is suppressed by
\be
\bra O \ls t , x _1\rs  O \ls t,  x_2 \rs \ket \sim  t^{-\fr{4 \D}{3 (1+ w)} }  ,
\ee
where $\D$ means a conformal dimension of an operator $O$. We also check this result with the exact numerical simulation.

The rest of this paper is organized as follows. In section 2, after briefly reviewing the braneworld cosmology in a five-dimensional $p$-brane gas geometry, we investigate mutual information in expanding universes, which describes a collective correlation between two macroscopic subregions. In section 3, we take into account two-point functions of local operators in the FLRW universes expanding exponentially or by a power law. We also find the analytic form of the suppression powers by applying the holographic technique. Lastly, we finish this work with some concluding remarks in section 4.


\section{FLRW cosmology in the braneworld model }

To know the history of our universe, it is important to measure various cosmological data such as correlation functions. In general, understanding those cosmological data is challenging for expanding universes because the variation of spacetimes makes equations nontrivial. To detour this difficulty in this work, we take into account a holographic model realizing the standard FLRW cosmology and investigate various correlation functions in expanding universes. Before discussing the holographic model, we first summarize important features of the standard cosmology for later comparison with the holographic model. When our universe is filled with one kind of matter, the scale factor is determined by the Friedmann equation \ct{kinney2009tasi}
\be 		
\ls \fr{\dot{a}}{a} \rs^2 =  \fr{\k^2}{3} \fr{\r}{a^{3(1+w)}}  ,   \la{equation:ScaleFactor}
\ee
where $\r$ and $w$ are the energy density and equation of state parameter of the matter, respectively. Solving the Friedmann equation, the scale factor is determined as the function of a cosmological time  
\be
a (t) \sim t^{\fr{2}{3(1+w)}}   .
\ee
If the matters filling in the universe are $n$-dimensional objects, the equation of state parameter is given by \ct{Park:2020xho, Kim:2023fbr, Park:2022abi}. 
\be
w = - \fr{n}{3}  .
\ee
For example, radiations and massive particles have $w=1/3$ with $n=-1$ and $w=0$ with $n=0$, respectively. On the other hand, cosmic strings result in $w=-1/3$ with $n=1$. If we consider three-dimensional objects, they fill in the space and give rise to a vacuum energy with $w=-1$.

To describe this standard cosmology in the holographic setup, we take into account the braneworld model in a $p$-brane gas geometry. Assuming that $p$-branes extending to the radial direction of  a five-dimensional AdS space are uniformly distributed, their gravitational backreaction leads to the following $p$-brane gas geometry \ct{Park:2020jio, Koh:2020rti, Park:2020xho, Park:2021wep}
\be		  \la{metric:ansatz}
ds^2  &=&  - g_{\ta \ta}  d \ta ^2  + g_{rr}  dr^2 + g_{ij}\  d x^i dx^j   \nn
&=& - \fr{r^2}{R^2}  f (r) \ d \ta^2 + \fr{R^2}{ r^2 f(r) } dr^2  + \fr{r^2  }{R^2} \ \d_{ij} d x^i dx^j  ,    
\ee 
where $\ta$ indicates a bulk time and the metric factor $f(r)$ is given by
\be
f(r) = 1  -  \fr{\r_p}{r^{4-p}}  ,
\ee
where $\r_p$ means a number density of $p$-branes. We assume that two five-dimensional $p$-brane gas geometries are bordered at a four-dimensional hypersurface, which we call a braneworld (see more details in \cite{Park:2020jio, Park:2020xho}). In this case, the radial motion of the braneworld is governed by the junction equation  
\be
\pi^{(+)}_{\m\n} - \pi^{(-)}_{\m\n} = T_{\m\n} ,
\ee
where $(\pm)$ indicate two $p$-brane gas geometries on the both sides of the braneworld. Imposing a $Z_2$ symmetry for convenience, the canonical momenta of the boundary metric $\g_{\m\n}$ are given by
\be
\pi^{(+)}_{\m\n} = - \pi^{(-)}_{\m\n} = \fr{ 1}{2 \k^2}  \ls K_{\m\n}- \g_{\m\n} K \rs ,
\ee
where $K_{\m\n} = \nabla_\m n_\n$ is an extrinsic curvature and $n_\n$ is a unit normal vector.  Parameterizing an energy-momentum tensor of the braneworld as
\be
T_{\m\n} =  \fr{\s}{2 \k^2} \ \g_{\m\n} ,
\ee
the junction equation in the spatial section is reduced to
\be
K_{ij} - \g_{ij} K =  \fr{\s}{2}  \  \g_{ij}   .  \la{Equation:Junction}
\ee

Defining a cosmological time $t$ in the braneworld as
\be
d t^2 = g_{\ta \ta}  d \ta^2 - g_{rr}  dr^2 =   \lb g_{\ta \ta}  - g_{rr}  \ls \fr{dr}{d \ta} \rs^2 \rb  d \ta^2  ,    \la{Relation:cosmotime}
\ee 
the induced metric in the braneworld becomes
\be		 
ds^2_B = - d t^2 + \fr{r(t)^2}{R^2}  \  \d_{ij} d x^i dx^j  ,
\ee
where $r(t)$ means the radial position of the braneworld. This is the FLRW-type metric with the following scale factor 
\be
a(t) = \fr{r(t)}{R} , \la{Result:scalefactorr}
\ee 
where the radial position of the braneworld is reinterpreted as the scale factor of the braneworld. Rewriting the previous junction equation in terms of the cosmological time, we obtain
\be		
\dot{r}^2  = \fr{\s^2 \,  r^2  }{36} -  \fr{1}{g_{rr}} 
= \fr{ ( \s^2 - \s_c^2 ) \, r^2  }{36}   +  \fr{\r_p}{ R^2 } \fr{1}{r^{2-p}} ,
\ee
with a critical tension  $\s_c = 6/R$, where the dot means a derivative with respect to the cosmological time. As a result, the junction equation in the braneworld model plays a role of the Friedmann equation.

\begin{itemize}

\item For $\s \ne \s_c$ with $\r_p = 0$ or $\s = \s_c$ with $\r_4 \ne 0$, the junction equation is reduced to
\be
\ls \fr{\dot{r} }{r} \rs^2  = \fr{ \s^2 - \s_c^2 }{36} \quad  {\rm or} \quad  \fr{\r_4}{R^2} ,   \la{Equation:junction4}
\ee
where the right-hand side can be regarded as a cosmological constant. Solving this equation, the braneworld position becomes 
\be
r (t)  = R \, e^{H t}  ,		\la{Result:branetime}
\ee
where a Hubble constant is defined as $H = \sqrt{\s^2 - \s_c^2}/6$ or $\sqrt{\r_4}/R$. This solution describes an eternal inflation with $w=-1$.

\item For $\s = \s_c$ with $p \ne 4$,  the junction equation reduces to  
\be		    	\la{result:FriedmannonBrane}
\ls \fr{\dot{r} }{r} \rs^2 
 =   \fr{\r_p}{ R^2} \fr{1}{r^{4-p}}   .   \la{Result:juneq}
\ee
Since an observer living in the braneworld cannot see the radial direction, the observer measures the bulk $p$-branes as $(p-1)$-dimensional objects. In the braneworld model, therefore, the bulk $p$-branes are  identified with $n$-dimensional matters of the standard cosmology   
\be
w= - \fr{p-1}{3} = - \fr{n}{3} .    \la{Relation:EoSP}
\ee 
Then, the junction equation in \eq{Result:juneq} is equivalent to the Friedmann equation in \eq{equation:ScaleFactor}.
This shows that the braneworld model can realize the standard cosmology correctly. In the braneworld model, as a consequence, the scale factor results in
\be
a (t) =  a (t_0)  +  \fr{(2-p/2)^{\frac{2}{4-p}}  \, \rho _p^{\frac{1}{4-p}} }{  R^{ \frac{6-p}{4-p} } }  \,  \left(t-t_0\right){}^{\frac{2}{4-p }} ,
\ee
where $t_0$ is an appropriate initial time.
 
\end{itemize}

In the holographic setup, unlike the standard cosmology, one can easily calculate various correlation functions in expanding universes. For example, the time-dependent entanglement entropy in expanding universes has been studied in the boundary dS model \ct{Chu:2016uwi, Chu:2016pea, Koh:2018rsw} and the braneworld model \ct{Park:2020jio, Park:2021wep}. However, the boundary dS model cannot be applied to the other FLRW cosmologies expanding by power-law because their dual gravity theories still need to be discovered. Unlike the boundary dS model, the braneworld model can describe the other FLRW cosmologies, as shown above. Due to this reason, the entanglement entropy of the FLRW cosmologies has been studied in the braneworld model \ct{Park:2020jio, Park:2021wep}. It was shown that the entanglement entropy in the late time era increases by \ct{Park:2020jio}
\be
S_E \sim t^{\fr{4}{3 (1+w)}} .      \la{Result:SEforMI}
\ee 
Going beyond the entanglement entropy, this work further investigates mutual information between two macroscopic disjoint regions and two-point functions of local operators or excitation in expanding universes.

\begin{figure}
	\begin{center}
		\vspace{-2cm}
		\hspace{-0.5 cm}
		\subfigure{ \includegraphics[angle=0,width=0.5 \textwidth]{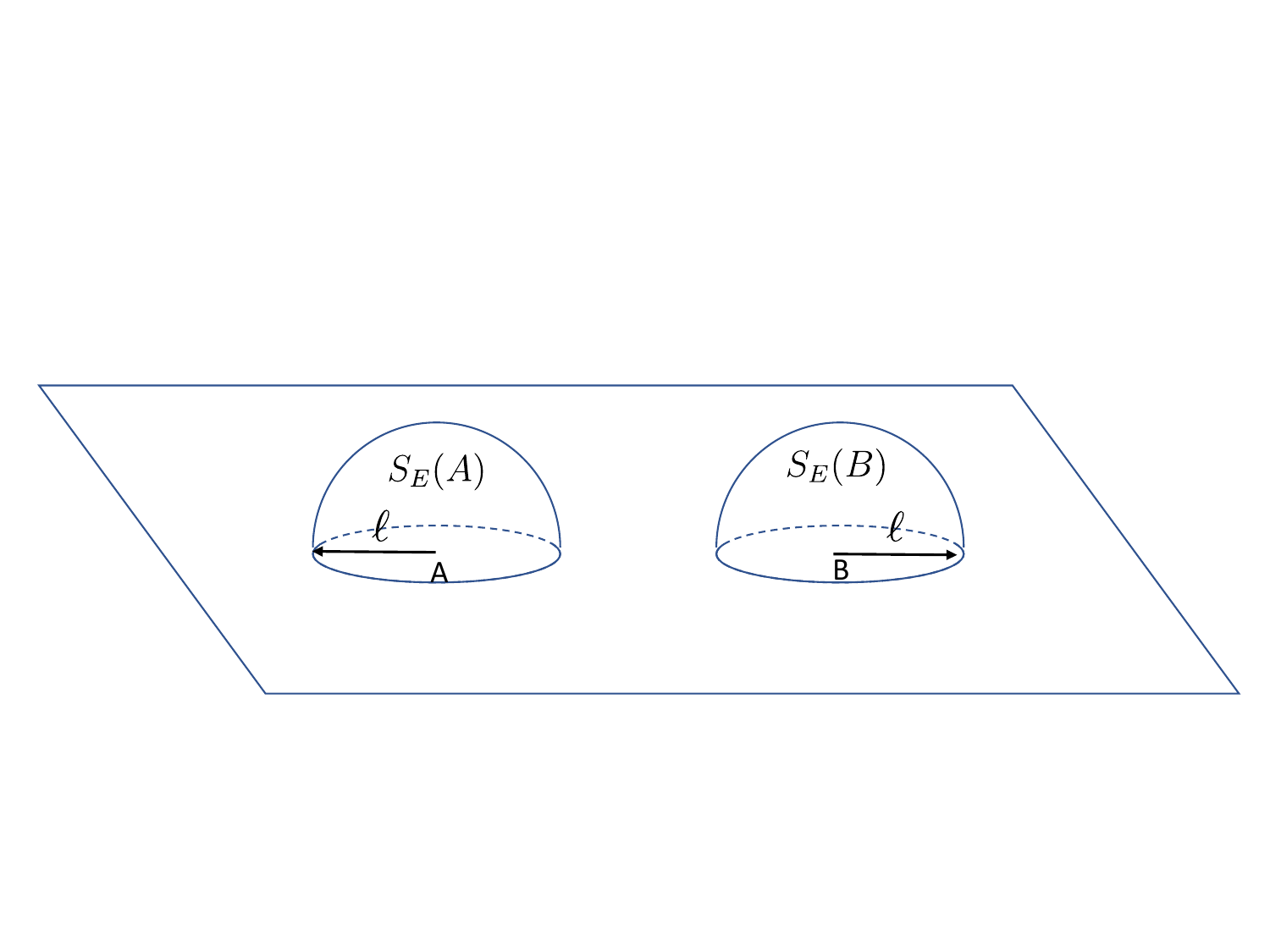}}
		\hspace{-0.5 cm}
		\subfigure{ \includegraphics[angle=0,width=0.5 \textwidth]{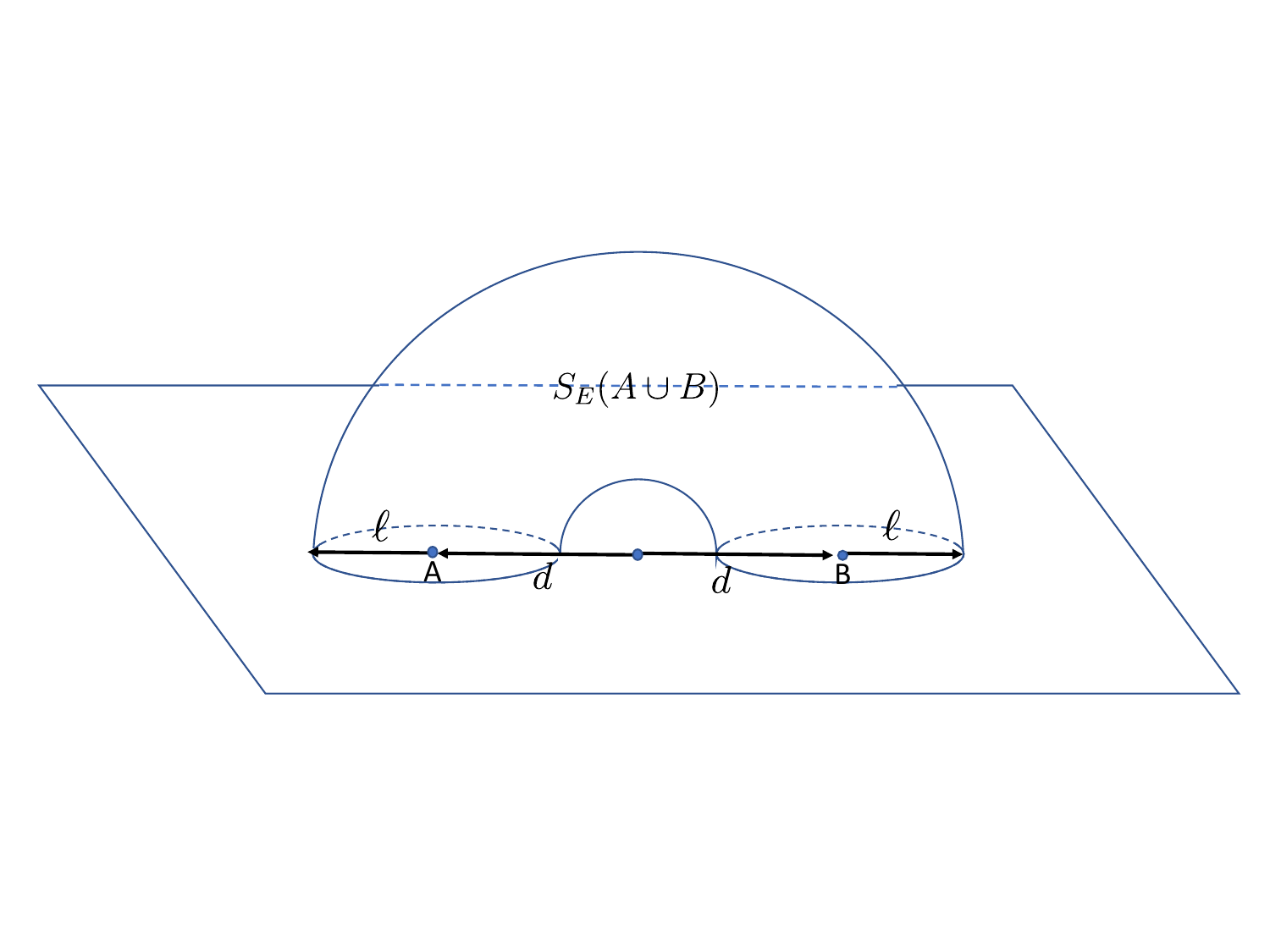}}
		\vspace{-1.5cm}
		\caption{We depict entanglement entropies, $S_E (A)$ and $S_E (B)$, of two disjoint regions  (left) and the entanglement entropy $S_E (A \cup B)$ of $A \cup B$ (right). Here, we assume that two disjoint regions with the same size $\bar{\r}$ have a distance $2 d$. }
	\end{center}
\end{figure}
 
We first look into mutual information between two disjoint regions, as shown in figure 1. In general, mutual information describes the collective correlation of background matters in two macroscopic regions, $A$ and $B$. The mutual information is defined in terms of the entanglement entropy \ct{Headrick:2010zt, Wolf:2007tdq, Molina-Vilaplana:2011ydi, Narayanan:2018ilr}
\be
MI (A,B) \equiv S_E (A) + S_E (B) - S_E (A \cup B)  \ge \fr{ \ls  \bra O_A \, O_B \ket - \bra O_A\ket \, \bra O_B \ket  \rs^2 }{2 |O_A|^2  \,  |O_B|^2  } \ge 0 ,
\ee
which must be positive. For convenience, we consider two macroscopic regions, $A$ and $B$, with the same size $\ell$ in the $p$-brane gas geometry
\be
ds^2 = \fr{R^2}{z^2} \ls - f(z) d \ta^2 + d x^2 + x^2 d \O_{2}^2 + \fr{1}{f(z)} \, dz^2 \rs , 
\label{Metric:pbraneGas}
\ee
where $x = | \vec{x} |$ and $z = R^2/r$. In the holographic setup, the entanglement entropy is determined as the area of a minimal surface anchored to the entangling surface. Then, the entanglement entropy is governed by
\be
S_E (A) = S_E (B) = \fr{  R^3}{ 4 G} \int d \O_2   \int_0^{\ell} d x \, \fr{x^2}{z^3 } \fr{\sqrt{f(z) + z'^2}}{\sqrt{f(z)}}  .   \la{Action:entangleAB}
\ee
From now on, we focus on mutual information in the late time era. In the late time era, the braneworld position approaches $z=0$, and $f(z)$ is approximated by $1$. Therefore, the configuration of a minimal surface is determined by the following equation of motion at leading order
\be
0 = z'' +\frac{2 z'^3}{x}+\frac{3 z'^2}{z}+\frac{2 z'}{x}+\frac{3}{z} ,
\ee
where the prime means a derivative with respect to $x$. Denoting the braneworld position as $\bar{z}$, which is in terms of the cosmological time 
\be
\bar{z} = \ls \fr{ 2 R^{5-p}}{ 4-p} \rs^{ \frac{2}{4-p}}   \ 
\rho _p^{-\frac{1}{4-p}}  \left(t-t_0\right)^{-\frac{2}{4-p}} ,
\ee 
the configuration of the minimal surface becomes at leading order 
\be
z (x) \approx \sqrt{\ell^2 + \bar{z}^2 - x^2} .
\ee
After plugging this solution into \eq{Action:entangleAB} and performing the integral, the entanglement entropy results in
\be
S_E (A) = S_E (B) \approx \fr{R^3 }{8 G} \fr{\ell^2 \O_2}{\bar{z}^2} =  4 \pi \ell^2  c_p   \,  \rho _p^{-\frac{2}{4-p}}  \,  t^{-\frac{4}{4-p}} ,
\ee 
with
\be
c_p =  \ls \fr{ 4 -p } {2 R^{ 5-p}}\rs^{ \frac{4}{4-p}},
\ee
which is the entanglement entropy \eq{Result:SEforMI} in expanding universes.

Now, we calculate the entanglement entropy of $A \cup B$. The entanglement entropy connecting $A$ and $B$, as shown in figure 1, is governed by
\be
S_E (A \cup B) = \fr{  R^3}{ 4 G} \int_0^{2 \pi}  d {\ph}  \int_0^\pi d \th \sin \th    \int_0^{d + \ell \cos \th} d x  \, \fr{x^2}{z^3 } \fr{\sqrt{f(z) + z'^2}}{\sqrt{f(z)}} .
\ee
This allows the following cylindrical configuration in the late time era
\be
z (x) \approx \sqrt{ (d + \ell \cos \th )^2 + \bar{z}^2 - x^2},
\ee
where $2 d$ means the distance between two disjoint regions. This solution results in
\be
S_E (A \cup B) = \fr{ R^3}{8 G } \fr{4 \pi (3 d^2 + \ell^2)  }{3 \bar{z}^2} =   \fr{4 \pi (3 d^2 + \ell^2)  }{3}  c_p  \,  \rho _p^{-\frac{2}{4-p}}  \,  t^{-\frac{4}{4-p}} ,
\ee
In the late time era, therefore, the mutual information becomes
\be
MI (A,B) \approx \fr{R^3 }{8 G} \fr{4 \pi  (5 \ell^2 - 3 d^2)}{ 3 \bar{z}^2} = \fr{ 4 \pi  (5 \ell^2 - 3 d^2)}{ 3 \bar{z}^2}  c_p \, \rho _p^{-\frac{2}{4-p}}  \,  t^{-\frac{4}{4-p}}  .
\ee
This result shows that the mutual information decreases with the matter density and time. Moreover, since the mutual information must be positive, there is no mutual information when the distance of two macroscopic regions is larger than the subsystem size, $d \ge \sqrt{5/3} \,  \ell$ at leading order.

\section{Microscopic correlations in expanding universes }

In the previous section, we showed that the braneworld model can realize the standard cosmology correctly and studied the mutual information between two macroscopic disjoint regions. From now on, we delve into microscopic correlations of a local operator or excitation in expanding universes. At the universe scale, stars may be regarded as local heavy operators. When stars were born, we may describe their correlations in terms of two-point functions in the expanding universe and understand their time dependence.

\subsection{Two-point function in an eternally inflating universe}

We first take into account an eternally inflating universe. When a universe has a positive cosmological constant, its background geometry is described by a dS space and results in eternal inflation. In this dS background geometry, the correlation function between two local operators with a conformal dimension $\D$ is given by
\be
\bra O \ls T_1, \vec{x}_1\rs  O \ls T_2,  \vec{x}_2 \rs \ket   \sim 
 \left(\frac{  T_1 T_2 }{ - | T_1 -  T_2 |{}^2+ | \vec{x}_1 - \vec{x}_2 |{}^2  }\right)^{\Delta } ,   
 \la{Result:2ptCfuntion}
\ee
where $T_i$ is a conformal time in the comoving frame. In the boundary dS model, it was shown that the same correlator can be rederived  \ct{Kim:2023fbr, Chu:2016uwi, Chu:2016pea}. Again, we exploit the braneworld model as the bulk geometry allowing the FLRW cosmology is not known yet for the boundary dS model. Using the braneworld model, we first study a two-point function in an eternally inflating universe and verify its consistence with the CFT result \eq{Result:2ptCfuntion} in a dS space.

To describe eternal inflation in the braneworld model, we take into account the case of $\s \ne \s_c$ with $\r_p=0$. Introducing a new radial coordinate $z=R^2/r$ as \eq{Metric:pbraneGas} for convenience, the metric of a five-dimensional AdS space is rewritten as
\be
ds^2 = \fr{R^2}{z^2} \ls - d \ta^2 + d \vec{x} \cdot d \vec{x} + dz^2 \rs  .   \la{Metric:AdS}
\ee
In terms of the new radial coordinate, the junction equation in \eq{Equation:junction4} is rewritten as 
\be
\fr{d z}{dt}   =  - H z  \quad {\rm with} \quad  H =  \fr{\s^2 - \s_c^2}{6} ,  \la{Relation:z-t}
\ee
where $t$ is the cosmological time in \eq{Relation:cosmotime}. Using the chain rule, the cosmological time $t$ is related to the bulk time $\ta$
\be
 \fr{ dt}{d \ta} = \fr{R}{z \, \sqrt{1 + H^2 R^2 }}  .  \la{Relation:t-tau}
\ee
Combining two relations in \eq{Relation:z-t} and \eq{Relation:t-tau} determines the relation between $z$ and $\ta$
\be
\fr{d \ta}{dz} =  -  \fr{ \sqrt{1 + H^2 R^2 }}{H R} .
\ee
Solving these relations, we obtain
\be
z (t) = z_0  \, e^{- H  (t- t_0)}    \quad {\rm and}  \quad
\ta (z) =  -  \fr{ \sqrt{1 + H^2 R^2 }}{H R}  z   ,    \la{Result:ztauitot}
\ee
where $t_0$ is introduced as an appropriate initial (or reference) time and $z_0$ corresponds to the radial position of the braneworld at $t=t_0$. 

To describe the dS space in the comoving frame, we introduce a conformal time 
\be
T =  \fr{z}{H R} = \fr{1}{H} \ e^{- H (t-t_0)}    .
\ee
In the comoving frame, then the induced metric on the braneworld becomes  
\be		 
ds^2_B =  \fr{1}{H^2  T^2} \ls - d T^2 +  \d_{ij} d x^i dx^j  \rs .   \la{Metric:dSspace}
\ee
This is exactly the same as a four-dimensional dS metric. Using the solutions in \eq{Result:ztauitot}, we can see that the conformal time satisfies the following relation 
\be
- dT^2 =  - d \ta^2 + d z^2 .    \la{Relation:bwtime}
\ee
We suppose that two local operators are located at $\lc \et_1, x_1 , z_1 \rc$ and $\lc \et_2,x_2 , z_2\rc$ where $z_i$ are the braneworld positions at given Euclidean time, $\et_i = i  \ta_i$, and $x_i = |\vec{x}_i|$. Since a geodesic curve connecting two operators is associated with a two-point function in the holographic model, a geodesic length in the bulk AdS space \eq{Metric:AdS} is governed by
\be
L (\et_1, x_1,z_1;\et_2,x_2, z_2)  = R \int_{\et_1}^{\et_2}  d \et  \ \fr{ \sqrt{ 1  + \dot{x}^2 + \dot{z}^2} }{z}  ,  \la{Action:dualCFT}
\ee
where the dot indicates a derivative with respect to $\et$. Regarding the geodesic length as an action, it has two conserved charges 
\be
P  
=  \fr{R \, v}{z_t \sqrt{ 1  + v^2 }}  
\quad {\rm and} \quad 
H 
=  - \fr{R}{z_t \sqrt{ 1  + v^2 }}   \la{Result:consevedH}  ,
\ee
where $z_t$ means a turning point at which $\dot{z}=0$ and $\dot{x} = v$

Comparing $P$ with $H$,  we find that $\dot{x} = v$ must be constant. After solving it, the operator's position is determined as a function of the bulk time 
\be
x (\et) = v \,  \et  + c_1  .
\ee
Imposing the boundary conditions, $x(\et_1) =x_1$ and  $x(\et_2) =x_2$, the integral constant and velocity are determined by
\be
c_1 = - v \et_1 +  x_1 \quad {\rm and} \quad  v = \fr{\D x}{\D\et }  ,
\ee
where $\D x = | x_1 - x_2|$ and $\D \et = | \et_1 - \et_2 | $. From \eq{Result:consevedH}, we also determine the relation between $z$ and $\et$
\be
\dot{z} =   \fr{  \sqrt{1+v^2} \, \sqrt{z_t^2 - z^2}}{z} .  \la{Equation:radialz}
\ee
Solving this equation with the boundary conditions, $z (\et_1) = z_1$ and  $z (\et_2) = z_2$, we arrive at
\be
z (\et) =  \sqrt{z_t^2 - (1 + v^2) \, (\et - c_2)^2}  .
\ee
with the following constants
\be
c_2 =  \et_1 + \fr{\sqrt{z_t^2 - z_1^2}}{\sqrt{1+v^2}}  \quad {\rm and} \quad
z_t = \frac{\sqrt{(1 + v^2) \, \D \et^2 + \D z^2 } \ \sqrt{ (1 + v^2) \, \D \et^2 + ( z_1 + z_2)^2 }   }{2 \sqrt{1 + v^2} \, \D \et} ,
\ee
where $\D z = | z_1 - z_2 |$. This solution describes the radial position and the scale factor of the braneworld in the bulk.

After plugging the obtained solutions into the geodesic formula, the minimal geodesic length connecting two operators at $\lc \et_1, x_1\rc$ and $\lc \et_2, x_2\rc$ is given by
\be
L(\et_1,x_1;\et_2,x_2) =   \ \ls  \int_{z_t}^{z_1} dz  -  \int_{z_2}^{z_t} dz \rs dz \fr{ R \, z_t}{z \sqrt{z_t^2 - z^2}}   ,
\ee
where $z_1$ and $z_2$ are determined as function of $\et_1, x_1, \et_2$ and $x_2$. After performing the integral and applying the inverse Wick rotation ($\et=i \ta$), we finally obtain the following minimal length
\be
L(T_1,x_1;T_2,x_2) = R \log \lb\fr{ \ls \sqrt{ - \D T^2 + \D x^2}  - \sqrt{- \D \ta^2 + \D x^2 + (z_1 + z_2)^2 }  \rs^2}{4 z_1 z_2} \rb  ,
\ee
where $\D T^2 = |T_1 - T_2|^2= \D \ta^2 -\D z^2$ due to \eq{Relation:bwtime}. If we specify the initial position of the braneworld at $t=t_0$ as
\be
z_0   = \frac{H R \left(x_1-x_2\right) e^{H (t_2 - t_0)}}{ \sqrt{ 1 - 2 \left(1 - 2 H^2 R^2\right) e^{H \left(t_2 - t_1 \right)} + \left(1-8 H^2 R^2\right) e^{2 H  ( t_2 - t_1)} }}  ,
\ee
the resulting two-point function, according to the holographic proposal in \eq{Formula:TPF}, results in
\be
\bra O \ls T_1, x_1\rs  O \ls T_2,  x_2 \rs \ket  
 =   \left(  \frac{ 4 H^2 R^2 \ T_1 T_2 }{ - | T_1 -  T_2 |{}^2+ | x_1 - x_2 |{}^2  }\right)^{\Delta } ,  
\ee
This is consistent with the CFT result \eq{Result:2ptCfuntion} in a dS space. At a given time with $T_1=T_2$, the two-point function in the inflating universe is reduced to
\be
\bra O \ls t, x_1 \rs  O \ls t,  x_2 \rs \ket  =   \ls \frac{ 2 R e^{H t_0} }{  | x_1 - x_2 |  }\rs^{2 \D} \ e^{- 2  \D  H  t }   . 
\ee
This shows that the two-point function exponentially suppresses in the inflating universe  because of the exponential expansion. In the expanding universe, this result also shows that the correlation function suppresses more rapidly as the conformal dimension becomes large. 


\subsection{Two-point functions in FLRW cosmologies}

Now, we investigate two-point functions in the other FLRW cosmologies expanding by power-law. To do so, we take into account the case of $p \ne 4$ with $\s = \s_c$. In this case, the $p$-brane gas geometry is written as
\be		 
ds^2 = \fr{R^2}{z^2}  \ls  -  f (z) \ d \ta^2 + \fr{1}{f(z) } dz^2  +  \d_{ij} \, d x^i dx^j  \rs , \la{Metric:pgasgeoz}
\ee 
with the following blackening factor
\be
f(z) = 1  -  \fr{z^{4-p}}{z_h^{4-p}}    ,
\ee
where the black hole horizon $z_h$ is related to the number of $p$-branes, $\r_p   = R^{2(4-p)} / z_h^{4-p} $. In the braneworld model, the braneworld position $\bar{r} = R^2/\bar{z}$ can be reinterpreted as the scale factor in \eq{Result:scalefactorr}, which in terms of $\bar{z}$ becomes
\be 
\bar{z} = \fr{R}{a (t) } \sim t^{- 2/ (4- p)} .    \la{Relation:brpositionscale}
\ee

To calculate a geodesic length, we assume that the braneworld is located at $\bar{z}$ at a given time $t$ and that two operators are located at $\lc t, x_1, \bar{z} \rc$ and $\lc t, x_2, \bar{z} \rc$. In the braneworld model, the time $t$ can be determined as a function of the braneworld position $\bar{z}$ due to \eq{Relation:brpositionscale}. Therefore, the geodesic length at a given time $t$ is governed by
\be
L \ls t, x_1 ; t,  x_2 \rs = R \int_{x_1}^{x_2} dx \fr{ \sqrt{f + z'^2}}{z \sqrt{f}} ,    \la{Action:geodlength}
\ee
Introducing a turning point $z_t$, the geodesic length extends to only the range of $\bar{z} \le z \le z_t$. 
Using this fact,  we determine the distance and geodesic length in terms of the turning point and the braneworld position
\be
| x_1 - x_2 | &=& \int_{\bar{z}}^{z_t} dz \fr{2 z}{ \sqrt{f} \ \sqrt{z_t^2 -z^2}} , \nn
L \ls t, x_1 ; t,  x_2 \rs &=& \int_{\bar{z}}^{z_t} dz \fr{2 R z_t}{z  \sqrt{f} \ \sqrt{z_t^2 -z^2}}    .\la{Equation:disleng}
\ee
It is worth noting that, since the background geometry \eq{Metric:pgasgeoz} is time-independent, the equation of motion derived from \eq{Action:geodlength} is also time-independent. However, the time-dependent braneworld position $\bar{z}$ requires a time-dependent boundary condition. Due to this fact, we can finally obtain a
time-dependent geodesic length.

In the short-distance limit ($| x_1 - x_2 | \to 0$ and $z_t / z_h \to 0$), since a geodesic curve extends only to the asymptotic region, the blackening factor can be approximated by $f \approx 1$. In this case, leading terms of the distance and geodesic length are given by
\be
| x_1 - x_2 | & \approx & 2 z_t + \cdots   , \nn
L \ls t, x_1 ; t,  x_2 \rs  &\approx& 2 R \log \fr{2 z_t}{\bar{z}}  + \cdots   .
\ee  
Applying the holographic formula \eq{Formula:TPF}, therefore, the leading contribution to a short-distance two-point function results in
\be
\bra O \ls t, x_1 \rs  O \ls t,  x_2 \rs \ket  \sim \fr{\bar{z}^{2 \D}}{| x_1 - x_2 |^{2 \D}} \sim  \fr{1}{| x_1 - x_2 |^{2 \D}}  \  t^{- 4 \D/ (4- p)} ,     \la{Result:2ptfunearly}
\ee 
where we use \eq{Relation:brpositionscale}. In the long-distance limit ($| x_1 - x_2 | \to \infty$ and $z_t / z _h \to 1$), on the other hand, the geodesic length  is rewritten as the following form
\be
L \ls t, x_1 ; t,  x_2 \rs  = \lim_{z_t \to z_h} \ls \fr{R | x_1 - x_2 |  }{z_t}  + 2  R \int_{\bar{z}}^{z_t} dz \fr{\sqrt{1 -  z^2/z_t^2}}{ z \sqrt{f}} \rs   .
\ee
In the late time era ($t \to \infty$ and $\bar{z} \to0$), the first term gives rise to a dominant contribution. Therefore, the geodesic length can be approximated by
\be
L \ls t, x_1 ; t,  x_2 \rs   = \fr{R | x_1 - x_2 |  }{z_h}   - 2 R \log \bar{z} + {\cal O}  \ls 1 \rs .
\ee 
Using this fact, the leading two-point function in the long-distance limit and the late time era is reduced to
\be
\bra O \ls t, x_1 \rs  O \ls t,  x_2 \rs \ket   
\sim  e^{-    \D  | x_1 - x_2 | / z_h }    \   t^{- 4 \D/ (4- p)}    .     \la{Result:2ptfunlate}
\ee
Therefore, when one kind of matter with the equation of state parameter $w$ is uniformly distributed, the holographic study shows that a short-distance two-point function in the FLRW cosmology becomes 
\be
\bra O \ls t, x_1 \rs  O \ls t,  x_2 \rs \ket   
\sim  \fr{1 }{| x_1 - x_2 |^{2 \D}}  \  t^{- \fr{4 \, \D }{3 (1 +  w)} }  ,   \la{Result:2ptfunFLRW1}
\ee
whereas a long-distance two-point function reads
\be
\bra O \ls t, x_1 \rs  O \ls t,  x_2 \rs \ket   
\sim  e^{-    \D  | x_1 - x_2 | / z_h }  \   t^{- \fr{4 \, \D }{3 (1 +  w)} }  .   \la{Result:2ptfunFLRW2}
\ee

In general, the mass of a matter leads to a relevant deformation and breaks the conformal symmetry. In the short-distance limit, the effect of a massive matter becomes subdominant. This is why the short-distance two-point function in the spatial direction is suppressed by a power-law. In the long-distance limit, however, the spatial two-point function decays exponentially due to the screening effect caused by the interaction with the background matter. This screening effect can be reinterpreted as the effective mass of the local operator in the medium 
\be
m_{eff} = \fr{1}{\xi}
= \fr{\D }{R^{2 }} \ \r_p^{\fr{1}{3 (1+w)}}  ,
\ee
where $\xi$ means a correlation length. In the braneworld model, this implies that the two-point function decreases by a power law in the short-distance limit ($ | x_1 - x_2 |  < \xi$), while it exponentially suppresses in the long-distance limit ($ | x_1 - x_2 |  > \xi$). If we reinterpret the correlation length as a typical size of the correlation function, it decreases as the density of the matter increases, as expected. Moreover, it shows that the large $w$ or small $p$ gives rise to the large correlation length. This feature is similar to the previous macroscopic correlation functions called mutual information.  

In the braneworld model, $0$- and $1$-branes in the bulk map to radiations and massive particles in the braneworld, respectively. In the expanding universe with uniformly distributed radiations or massive particles, the time dependence of the two-point function can be determined by \eq{Equation:disleng}. Although we cannot perform the integrals analytically for a general $p$, it is possible to figure out a two-point function numerically. In Fig. 2, we depict (a) the time evolution of a two-point function with a fixed distance, $|x_1 - x_2| =1$, 
\be
\bra O \ls t , x _1\rs  O \ls t,  x_2 \rs \ket \sim  t^{-\g}  ,
\ee
and (b) the change of a suppression power with time. In the late time era, we see that the suppression powers in the numerical simulation are perfectly matched to the previous analytic expectation in \eq{Result:2ptfunFLRW2} 
\be
\g = \lp -   \fr{d \log \bra O \ls t , x _1\rs  O \ls t,  x_2 \rs \ket }{d \log t} \right|_{|x_1 - x_2| =1}= \fr{4 \D}{3 (1+ w)} .
\ee
We also check this result by the analytic calculation for $p=2$ (see Appendix A).

\begin{figure}
	\begin{center}
		\vspace{-0.5 cm}
		\hspace{-0.5 cm}
		\subfigure[Two-point functions]{ \includegraphics[angle=0,width=0.5 \textwidth]{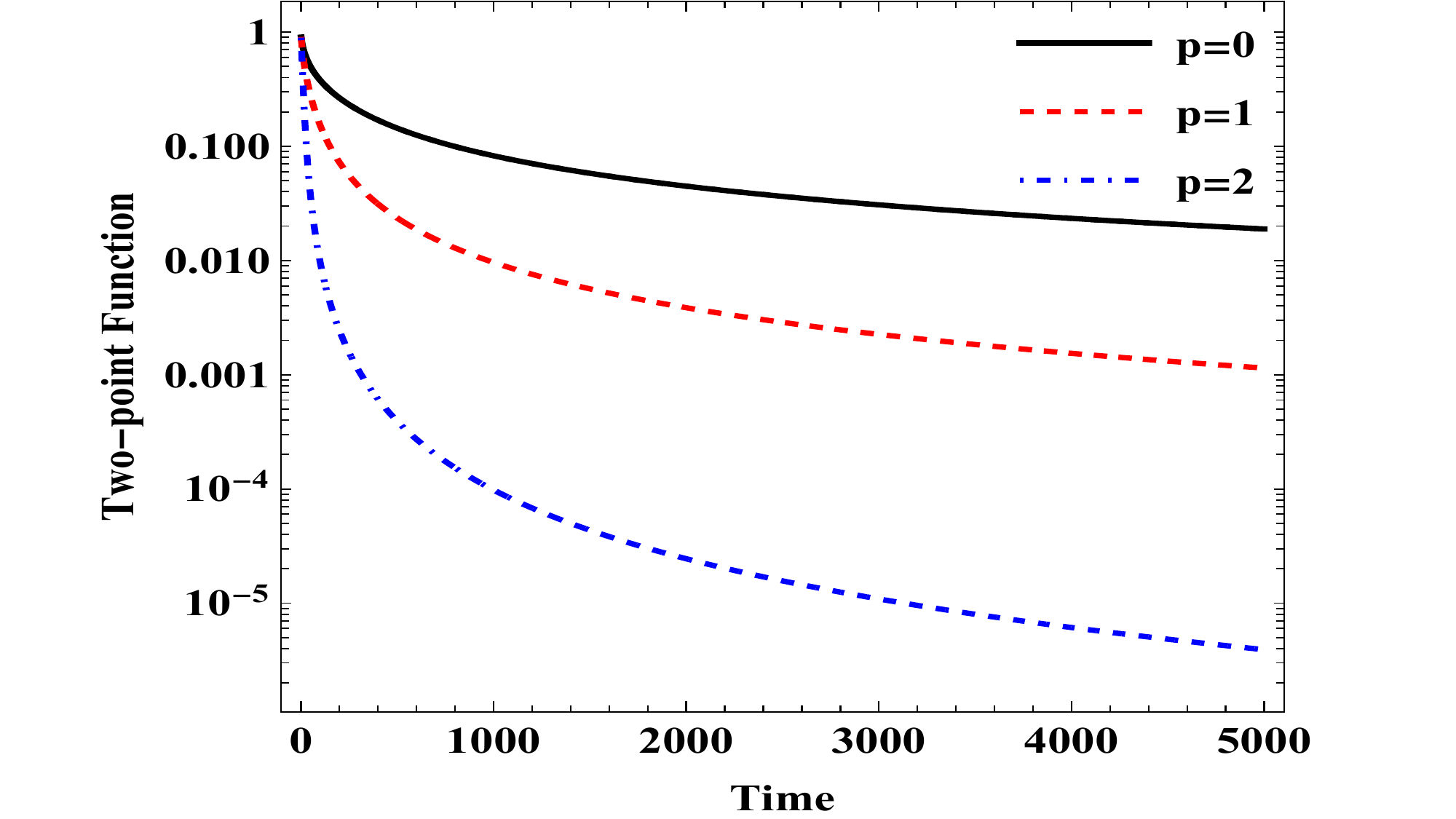}}
		\hspace{-0.5cm}
		\subfigure[Suppression power]{ \includegraphics[angle=0,width=0.5 \textwidth]{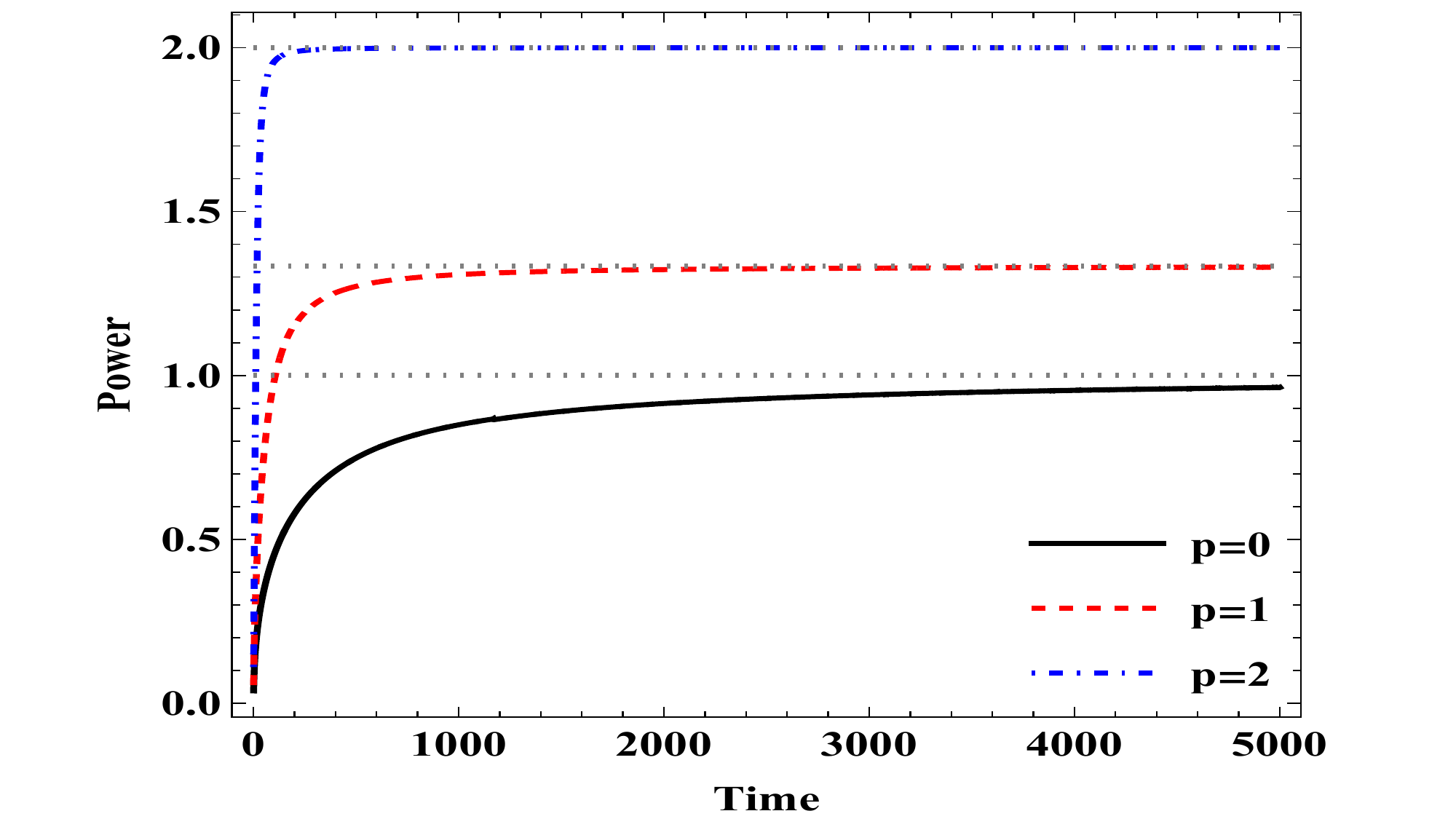}}
		\vspace{-0cm}
		\caption{ (a) Time evolution of a two-point function in FLRW cosmologies where we take $\D = |x_1 - x_2| = R  =1$ and $z_h = 10$. (b) In the late time era, we numerically obtain $\g = 1$ for $w=0$, $\g = 4/3$ for $w=-1/3$, and $\g = 2$ for $w=-2/3$.}
	\end{center}
\end{figure}


\section{Discussion}

In this work, we have investigated microscopic and macroscopic correlation functions in expanding universes. To do so, we took into account the braneworld model. When we consider the junction equation in the $p$-brane gas geometry, its solution results in the standard cosmology depending on the equation state parameter of the matter. In the braneworld model, the FLRW-type cosmology is described by the radial motion of the braneworld in the bulk geometry. In this case, $p$-branes in bulk are detected as $(p-1)$-dimensional objects to an observer living in the braneworld because the observer cannot see the bulk's radial direction. Therefore, the junction equation in the $p$-brane gas geometry determines the cosmology of the braneworld with uniformly distributed $(p-1)$-dimensional objects. We showed that the braneworld cosmology is consistent with the standard cosmology. Using the equivalence between the braneworld and standard cosmologies, we investigated time-dependent mutual information and two-point functions in the expanding universes by using the braneworld model.

We first improved the entanglement entropy calculation in \ct{Park:2020jio, Park:2021wep} into the time-dependent mutual information, which measures the entanglement between two disjoint macroscopic subregions in the expanding universes. We found that the mutual information decreases by a power law in the expanding universe
\be
MI (A,B) \sim  (5 \ell^2 - 3 d^2) \  t^{-\frac{4}{3 (1 +  w)}}  ,
\ee
where the suppression power decreases as the equation of state parameter $w$ increases. We also found that the mutual information becomes zero when the distance $d$ between two subregions is larger than a critical value $d_c$ determined by the subsystem size $\ell$
\be
d_c = \sqrt{\fr{5}{3} } \ \ell .
\ee 
This implies that there is no mutual information between two disjoint macroscopic regions if their distance is larger than the critical size, $d \ge d_c$. 

In the holographic setup, a two-point function of QFT is described by a geodesic curve extending to the dual geometry. To verify the computation of the two-point function in the braneworld model, we first calculated the two-point function in the inflating universe and successfully reproduced the known two-point function in a dS space. In the eternally inflating universe, since the metric relies on time explicitly, the time translation symmetry is broken. In the early time era, the spatial two-point function in the inflating universe is suppressed by a power-law. In the late time era, however, it is exponentially suppressed due to the exponential expansion in the inflation era
\be
\bra O \ls t, x_1 \rs  O \ls t,  x_2 \rs \ket  \sim    e^{- 2  \D  H  t }   . 
\ee

Next, we investigated the spatial two-point functions in the FLRW cosmology expanding by a power-law. In the short-distance limit, the spatial two-point function decreases by a power-law because the finite density effect is negligible in the UV region. We further showed that in the late time era, the two-point function in the FLRW cosmology is suppressed by a power law, unlike the previous inflating universe, 
\be
\bra O \ls t , x _1\rs  O \ls t,  x_2 \rs \ket \sim  t^{ - 4 \D / 3 (1+ w) }  ,   \la{Result:latetime2}
\ee
where the equation of state parameter $w$ is directly related to $p$
\be
 w = - \fr{p-1}{3} .
\ee
To validate this result, we first took into account the case of $p=2$ in the Appendix because it allows analytic calculation. We successfully rederived the above time-dependent two-point function. For $p=0$ and $p=1$ cases where the analytic calculation is unavailable, we performed numerical simulations for the time evolution of two-point functions. We found that the suppression powers in the numerical simulation are perfectly matched to the theoretical prediction in \eq{Result:latetime2}. In this work, we investigated two-point functions in terms of the distance between two local operators. It would be interesting to look into further the Fourier transformation of the obtained two-point functions because they give us more information about the power spectrum in the FLRW cosmology. We hope to report more results on this issue in future works.

\vspace{0.5cm}

{\bf Acknowledgement}

This work was supported by the National Research Foundation of Korea(NRF) grant funded by the Korean government(MSIT) (No. NRF-2019R1A2C1006639).


\appendix
\section{Exact microscopic two-point function for $p=2$}

To check the validity of the results in \eq{Result:2ptfunFLRW1} and \eq{Result:2ptfunFLRW2}, we focus on the case of $p=2$  because it allows an analytic calculation. To an observer living in the brane world, bulk $2$-branes are observed as one-dimensional cosmic strings with the equation of state parameter, $w=-1/3$. The $2$-brane gas geometry has the following metric factor
\be
f(z) =  1 -   \fr{z^2}{z_h^2} ,
\ee
where $\r_2 = R^4/z_h^2$, so the radial motion of the braneworld is governed by the following junction equation
\be		    
\dot{z}   =   - \fr{z^2}{ R z_h}   .
\ee
Solving this junction equation, the braneworld's radial motion is described by
\be
z =  c+  \fr{ R \, z_h}{t}  .     \la{Solution:zp2}
\ee
Requiring that $z \to 0$ at $t \to \infty$, the integral constant is fixed to be $c=0$. This solution implies that the braneworld linearly expands in time, $a(t) \sim 1/z \sim t$.

Now, we look into a spatial two-point function in the linearly expanding universe. We assume that two local operators are located at $\lc t, x_1 \rc$ and $\lc t,  x_2\rc$. Then, a geodesic curve connecting these two operators is governed by
\be
L \ls t , x_1 ; t ,  x_2 \rs = R \int_{x_1}^{x_2} dx \fr{ \sqrt{f + z'^2}}{z \sqrt{f}} ,   \la{Formula:gedp2}
\ee
where the prime means a derivative with respect to $x$. Using the translation symmetry in the $x$-direction, $z$ is given by a function of $x$
\be
\fr{dz}{dx} = \fr{\sqrt{f} \, \sqrt{z_t^2 - z^2} }{ z} .
\ee
A solution of this differential equation becomes
 \be
 x = x_t - z_h \tanh^{-1} \ls \fr{\sqrt{z_t^2 - z^2}}{\sqrt{z_h^2 - z^2}} \rs ,
 \ee
where $x_t$ is an integral constant. Since the geodesic formula in \eq{Formula:gedp2}  is invariant under $x \to x' =  (x_1 + x_2) - x$, a turning point must appear at $x = (x_1 + x_2)/2$ and $z=z_t$. This fact determines the integral constant as
\be
x_t= \fr{x_1 + x_2}{2} .
\ee

Denoting the braneworld position as $\bar{z}$, a geodesic curve must be anchored to the operator's position, $\lc t, x_1 \rc$ and $\lc t, x_2 \rc$ at $z=\bar{z}$. In this case, it should be noted that, since we consider two operators at the same time $t$, they have to lie in the same braneworld at $\bar{z}= z (t)$. Imposing this constraint, the turning point $z_t$ is determined as
\be
z_t = \fr{\sqrt{\bar{z}^2 + z_h^2  \sinh^2 (|x_1-x_2|/2 z_h) }}{\cosh  (|x_1-x_2|/2 z_h)} .
\ee
Substituting these solutions into the geodesic formula \eq{Formula:gedp2}, we finally obtain the following spatial two-point function in the linearly expanding universe  
\be
\bra O \ls t , x_1\rs \ O \ls t ,  x_2 \rs \ket  =  \ls  \fr{  \sqrt{R^2  + t^2     \sinh^2 (|x_1-x_2|/2 z_h) } - t    \sinh (|x_1-x_2|/2 z_h)  }{\sqrt{R^2 + t^2   \sinh^2 (|x_1-x_2|/2 z_h) } +  t   \sinh (|x_1-x_2|/2 z_h) }  \rs^{ \D}   .
\ee

In the late time era ($t \gg   R/ \sinh  (|x_1-x_2|/2 z_h)$), a short-distance two-point function for $|x_1-x_2|  \ll z_h \sim 1/T_H$ is described by
\be
\bra O \ls t , x _1\rs  O \ls t,  x_2 \rs \ket  \sim  \fr{1}{ |x_1-x_2|^{2 \D} }  \ t^{-2 \D}   .
\ee
On the other hand, a long-distance two-point function for $|x_1-x_2|  \gg 1/T_H$ reduces to
\be
\bra O \ls t , x _1\rs  O \ls t,  x_2 \rs \ket  \sim   e^{- \D |x_1-x_2| /z_h}  \  t^{-2 \D}  .
\ee
These results are consistent with the expectations in \eq{Result:2ptfunFLRW1} and \eq{Result:2ptfunFLRW2} for $p=2$ with $w=-1/3$.



%

\end{document}